\documentclass[%
 preprint,onecolumn,
 amsmath,amssymb,
 aps,
]{revtex4-1}
\usepackage{graphicx,color}
\usepackage{dcolumn}
\usepackage{bm}

\begin{document}


\title{Quest for potentials in the quintessence scenario\\
}
\author{Tetsuya Hara} 
\affiliation{Department of Physics, Kyoto Sangyo University, Kyoto 603-8555, Japan}
\email{hara@cc.kyoto-su.ac.jp}


\begin{abstract}
The time variation of the equation of state $w$ for quintessence scenario with a scalar field as dark energy is studied up to the third derivative ($d^3w/da^3$) 
with respect to the scale factor $a$, in order to predict the future observations and specify the scalar potential parameters with the observables.
The third derivative of $w$ for general potential $V$ is derived and applied to several types of potentials.  
They are the inverse power-law ($V=M^{4+\alpha}/Q^{\alpha}$), the exponential ($V=M^4\exp{(\beta M/Q)}$),
the cosine ($V=M^4(\cos (Q/f)+1)$) and the Gaussian types ($V=M^4\exp(-Q^2/\sigma^2)$), which are prototypical potentials for the freezing and thawing models.  
 If the parameter number for a potential form is $ n$, 
it is necessary to find at least for $n+2$ independent observations to identify the potential form and the evolution of the scalar field ($Q$ and $ \dot{Q} $). 
 Such observations would be the values of $ \Omega_Q, w, dw/da. \cdots $, and $ dw^n/da^n$. 
 Since four of the above mentioned potentials have two parameters, it is necessary to calculate the third derivative of $w$ for them to estimate the predict values.   
  If they are tested observationally, it will be understood whether the dark energy could be described by the scalar field with this potential.  
  Numerical analysis for $d^3w/da^3$ are made under some specified parameters in the investigated potentials.  
  It becomes possible to distinguish the freezing and thawing modes by the accurate observing $dw/da$ and $d^2w/da^2$ in some parameters.
\end{abstract}	

\maketitle

\section{Introduction}
\hspace{0.5cm} There are mainly two theoretical viewpoints to explain the accelerated universe.  One is related to modification of gravity and 
the other is associated with vacuum energy and/or matter field theories.  
Taking the latter viewpoint, we investigate the scalar fields in quintessence scenario how relevant it to the dark energy.

  In this scenario, the potential of the scalar field has $ n$ independent 
  parameters, so we recognize that in principle $n$ time derivatives of the equation of state with observable $\Omega_Q$ and $w$ are enough to specify the scalar potentials
  and to predict the higher derivatives.  In the paper \cite{1}, we have calculated the third derivative of the equation of state 
  for five scalar potentials to identify the models and to predict the future observations.  
  The first and second derivatives have been investigated in the paper \cite{2}.

Usually, the variation of the equation of state $w$ for the dark energy is described by \cite{4}
\vspace{0.1cm}
\begin{align}
 w(a)=w_0+w_a(1-a) , \label{act0}
\end{align}
where  $a, w_0, $ and $ w_a$ are the scale factor ($a=1$ at current), the current value of $ w(a) $ and the first derivative of $ w(a)$ by $w_a=-dw/da$, respectively.

  We have extended the parameter space, in the paper \cite{1}, 
\begin{align}
w(a)=w_0+w_a(1-a)+\frac{1}{2}w_{a2}(1-a)^2 +\frac{1}{3!}w_{a3}(1-a)^3 , \label{act1}
\end{align}
where $w_{a2}=-d^2 w/da^2$ and $w_{a3}=-d^3 w/da^3$.

\begin{figure}[ht]
\includegraphics[clip,width=12cm,height=6.5cm]{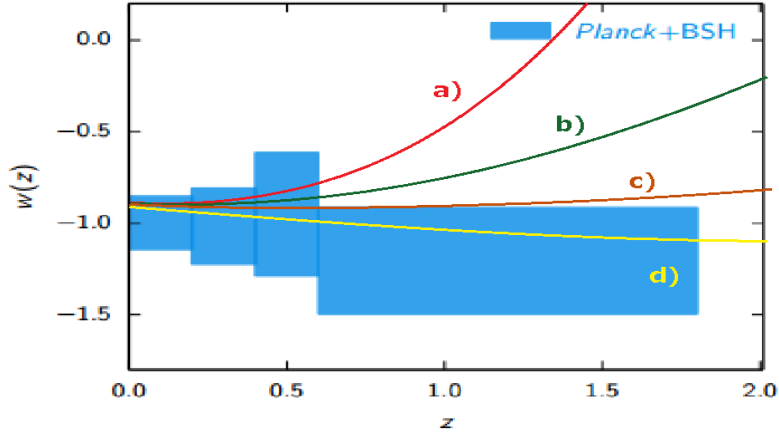}
\caption{From the observations, the reconstructed equation of state $w(z)$ as a function of red shift $z$ where $1+z=1/a $ \cite{3}. The simulated curves in Table 1 are designated by signs. }   
\end{figure}

Recent Planck and other observations for $w(z)$ are shown in Fig. 1 \cite{3}.  
The typical values for each $z$ are adopted and the derived values of $dw/da$ and $d^2w/da^2$ are estimated in Table 1.

\vspace{0.32cm}
\begin{table}
\begin{tabular}{|r|r|r|r|r|r|r|} \hline 
  \rule[0.2cm]{0.0cm}{0.2cm}   \hspace{0.5cm} \hspace{0.01cm} &   \hspace{0.3cm}$z=0.1$ \hspace{0.01cm}
 & \hspace{0.3cm}$ z=0.3 $\hspace{0.01cm} & \hspace{0.3cm} $z=0.5$ \hspace{0.01cm} & \hspace{0.3cm} $dw/da$ & \hspace{0.01cm} $d^2w/da^2$ \hspace{0.01cm}
  &  \hspace{0.1cm} sign \hspace{0.01cm}\\  [4pt] \hline 
 \rule[0.2cm]{0.0cm}{0.2cm}  
  \hspace{0.4cm}  $w(z) $  & \hspace{0.1cm} -0.90  & \hspace{0.1cm}  -0.91 &  -0.85 &  0.45 & 5.42  & a) \\  
  \hspace{0.5cm}  $w(z) $  & \hspace{0.1cm} -0.90  & \hspace{0.1cm}  -0.91 &  -0.90 &  0.17 & 1.39 & b) \\  
  \hspace{0.5cm}  $w(z) $  & \hspace{0.1cm} -0.90  & \hspace{0.1cm}  -0.93 &  -0.95 &  0.23 &  0.16  & c) \\  
  \hspace{0.5cm}  $w(z) $  & \hspace{0.1cm} -0.90  & \hspace{0.1cm}  -0.93 &  -0.958 &  0.18 & -0.48   & d) \\   
    \hline
  \end{tabular}
\vspace{0.4cm}
\caption{The values of $w(z)$ for each $z$ are adopted and the values $dw/da$ and $dw^2/da^2$ are estimated.  
The typical values are designated by signs, which are shown in Fig. 1 and plotted in Fig.2}   
\vspace{0.6cm}
\end{table}

We follow the single scalar field formalism of Steinhardt {\it et al.} (1999) \cite{5, 6} and investigate three potentials 
for so-called freezing model \cite {7}, in which the field is rolling towards down its potential minimum,
as $ V=M^{4+\alpha}/Q^{\alpha}$ (inverse power law) \cite{8}, $ V=M^4\exp(\beta M/Q)$ (exponential), and $V=M^{4+\gamma}/Q^{\gamma}\exp (\zeta Q^2 /M_{pl}^2)$ (mixed) \cite{9}. 
In this freezing model, $w(z)$ approaches to -1.

We study other two potentials for so-called thawing model, in which the field is nearly constant at first and then starts to evolve slowly down the potential;
 $V=M^4(\cos (Q/f)+1))$ (cosine)  and $V=M^4\exp(-Q^2/\sigma^2)$ (Gaussian).  In this thawing model, $w(z)$ starts from -1 and increases later.

 Because four of the above mentioned potentials have two parameters, it is necessary to calculate the third derivative of $w$ for them 
 to estimate the predict values.   
 If they are the predicted one, it will be understood that the dark energy could be described by the scalar field with this potential.  
  At least it will satisfy the necessary conditions.   
  Numerical analysis are made for $d^3w/da^3$ under some specified parameters in the investigated potentials except mixed one which has three parameters \cite{1}.

  \section{Equation of state $w_Q$ by a scalar field}
 \hspace{0.5cm} For the dark energy, we consider a scalar field $Q({\bf x},t)$, where the action for this field in the gravitational field is described by 
 \begin{align}
S=\int d^4 x \sqrt{-g} \left[ -\frac{1}{16 \pi G}R+ \frac{1}{2}g^{\mu \nu} \partial _ \mu Q \partial _ \nu Q -V( Q ) \right] +S_M ,  \label{act2}
\end{align}
 where $S_M$ is the action of the matter field and $G$ is the gravitational constant, occasionally putting $G=1$ \ \cite{4} .
 Neglecting the coordinate dependence, the equation for $Q(t)$ becomes 
 \begin{align}
\ddot{Q} + 3H \dot{Q}+V'=0  , \label{Qfield}
\end{align}
where $H$ is the Hubble parameter, over-dot is the derivative with time, and $V'$ is the derivative with $Q$.  
The equation of state $w_Q$ due to the scalar field is described by 
\begin{align}
w_ Q \equiv \frac{p_ Q}{\rho _ Q}=\frac{ \frac{1}{2} \dot{Q}^2-V}{ \frac{1}{2} \dot{Q}^2+V} . \label{state}
\end{align}
We put $w_Q=-1+\Delta $ for the later convenience ($ 0< \Delta < 0.2$).

\section{Second, and Third derivative of $w_Q$}
\hspace{0.5cm}  The detailed calculations of the second, and third derivatives of $w_Q$ for potentials  are displayed in the paper \cite {1}.
The numerical calculations for the freezing and thawing models under limited parameters are analyzed there.

In Fig. 2, the curve for $\alpha =0$ is presented for the case of $V=M^{4+\alpha}/Q^{\alpha}$ with $\Delta=0.1$ by the red solid curve 
in the $dw_Q/da$ and $d^2w_Q/da^2$ coordinates.  The signature of $\alpha$ will change beyond the parabolic curve.  
We assume $\alpha >0 $, so that the upper part of the red curve is forbidden for this potential and the freezing type potentials as well. 
 The green (inner) dotted curve is the case of $V=M^{4}(\cos (Q/f)+1)$ with $\Delta=0.1$.  Upper part of the green dotted curve is allowed region for this potential.
The allowed region of the other thawing potential (Gaussian) is the upper part of the red curve.

The interesting point is that the forbidden regions for the freezing type potentials are allowed region for the thawing type potentials and the reverse is also true. 
It is possible to distinguish the potentials among each type due to the different predicted values of $d^3w_Q/da^3$ \cite{1}, 
however it is necessary to make accurate observations for the values of $dw_Q/da$, $d^2w_Q/da^2$, $d^3w_Q/da^3$ and other parameters such as $\Delta$.


\begin{figure}[ht]
\includegraphics[clip,width=15cm,height=7.5cm]{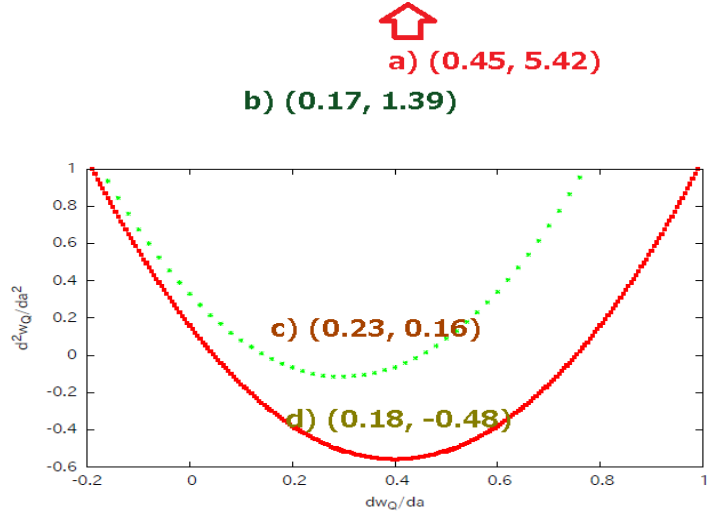}
\caption{The curve for $\alpha =0$ is presented for the case of $V=M^{4+\alpha}/Q^{\alpha}$ with $\Delta=0.1$ by the red solid curve in the $dw_Q/da$ and $d^2w_Q/da^2$ coordinates.
The signature of $\alpha$ will change beyond the parabolic curve and the upper part is forbidden for the freezing model.  The lower part of this curve is forbidden for the thawing model (Gausssian type). 
  The upper part of the green (inner) dotted curve is allowed region for thawing model of $V=M^{4}(\cos (Q/f)+1)$ with $\Delta=0.1$.  The typical values adopted in Table 1 are plotted. 
 Notice that the values of $d^2w_Q/da^2$ for a) and b) are out of frame.}   
\end{figure}

\section{Conclusion} 
\hspace{0.5cm} At present, backward observations, such as Planck, baryon acoustic oscillation, Supernova Ia, Hubble constant, weak lensing, 
and red shift distortion, have been undertaken to estimate $w_Q$ at the age $(1+z)$ as in Fig. 1 \cite{3}.
From Fig. 1, the rough values of $w(a=a_0)$, $dw/da$ and $d^2w/da^2$ have been estimated which are presented in Table 1. 
 They are pointed in the $dw_Q/da$ and $d^2w_Q/da^2$ plane in Fig. 2.

The adopted values from observation show $ -0.5< dw^2_Q/da^2 < 6$ within the region  $ 0.1 < dw_Q/da < 0.5 $.  
Although there is a lot of uncertainty, at the moment, it seems to be preferable for the thawing model against the freezing model under the comparison with the numerical results and the observations \cite {3}.

About observations in Fig. 1, it seems to be difficult to accept that the equation of state $w=p/\rho$ is almost smaller than -1 in the region $z \simeq 1 \sim 2$.  

Usually matter density increases as $\rho_m \propto (1+z)^3$, then $w=(p_Q+p_m)/(\rho_Q+ \rho_m)$ must increase with $z$ where $p_Q \simeq -\rho_Q $ and $ p_m \simeq 0$, 
taking that $p_Q$ and $p_m$ are pressure for scalar field and matter.  There seems to be no such features that $w$ increases with $z$ in the observation in Fig. 1.
 
If $w < -1$ which means $\Delta < 0$  is correct in Fig. 1, we must consider fully different models such as phantoms, 
quintom, k-essence, chameleon, tachyon, dilaton, modified gravity theory, and so on \cite{10}.


 


\end{document}